\documentclass[12pt]{article}
\usepackage{graphicx}
\usepackage{dcolumn} 
\usepackage{bm}      
\usepackage{umlaut}
\usepackage{epsfig}
\usepackage{dcolumn}
\usepackage{amsmath}
\usepackage{cite}
\usepackage{changebar}
\def\e{\epsilon}

\def\be{\begin{equation}}
\def\ee{\end{equation}}
\def\PL{{ Phys.\ Lett.\ }}
\def\PR{{ Phys.\ Rev.\ }}

\def\ZP{{ Z.\ Phys.\ }}
\def\EP{{ Europ.\ Phys.\ J.\ }}

\begin{document}

 10.6.2009 \hfill BI-TP 2009/12

\vskip2cm

\begin{center}
{\Large{\bf{The Speed of Sound in Hadronic Matter }}}

\vskip1cm

{\bf P.\ Castorina$^1$, J.\ Cleymans$^2$,
D.\ E.\ Miller$^{3,4}$ and H.\ Satz$^3$}
\\[1cm]
1 Dipartimento di Fisica, Universit{\`a} di Catania and
INFN Sezione di Catania\\
I-95123 Catania, Italy\\
2 Physics Department, University of Cape Town, South Africa,\\
3 Fakult\"at f\"ur Physik, Universit\"at Bielefeld, D-33501 Bielefeld, 
Germany\\
4 Department of Physics, Pennsylvania State University\\
Hazleton Campus,
Hazleton, PA 18202 USA\\
~\\

\bigskip

{\bf{Abstract:}
}
\end{center}

We calculate the speed of sound $c_s$ in an ideal gas of resonances, 
whose mass spectrum is assumed to have the Hagedorn form 
$\rho(m)\sim m^{-a}\exp\{bm\}$, which leads to singular behavior 
at the critical temperature $ T_c = 1/b $. With $ a = 4 $ the pressure 
and the energy density remain finite at $ T_c $, while the specific heat 
diverges there. As a function of the temperature, the corresponding speed 
of sound initially increases similarly to that of an ideal pion gas, 
until near $T_c$ resonance effects dominate, causing $c_s$ to vanish 
as $(T_c-T)^{1/4}$. In order to compare this result to the physical 
resonance gas models, we introduce an upper cut-off $M$ in the resonance 
mass integration. Although the truncated form still decreases somewhat 
in the region around $T_c$, the actual critical behavior in these models 
is no longer present.

\newpage

~~~~~~~~~~
\vskip1cm

\section{Introduction}

The abundant formation of resonances of increasing mass and rotational
degrees of freedom is one of the most striking features of strong interaction 
physics, which has attracted intense theoretical attention for half a 
century or more. Even before the quark infrastructure of hadrons was known, 
a self-similar composition scheme, the statistical bootstrap model (SBM), 
led to an exponentially increasing resonance spectrum \cite{Hage}. 
Shortly thereafter the dual resonance model (DRM) provided a description 
of hadron interactions in terms of the Regge resonance structure 
in the different kinematic channels \cite{Vene}, which again 
produced an exponentially increasing spectrum \cite{DRM}. 
In both these cases the basic feature is the underlying 
partition structure \cite{HW,S-ex,BFS}, leading to the form 
\be
\rho(m) \sim m^{-a} \exp{bm},
\label{spec}
\ee
for the number of states (degeneracy) of resonances of mass $m$.
This situation posed for the theory the two basic questions: what 
is the origin of so many resonances, and what is their effect on 
strong interaction thermodynamics?

\medskip

The first question was answered by the advent of the quark model and of
quantum chromodynamics (QCD), whereby the resonances are the different possible 
excitation states of the (initially) three quark flavors, together with their 
antiquarks. The second one was answered by R.\ Hagedorn \cite{Hage}, who 
showed that an exponentially rising resonance mass spectrum leads to
an upper bound on the temperature of hadronic matter, $T_c=1/b$,
with $b$ as specified in the eq.\ (\ref{spec}). Beyond a certain point 
an increase in the energy density of the medium does not increase 
its temperature; instead, more and higher mass resonances are formed. 
The behavior of the medium at the ``Hagedorn'' temperature $T_c$ is
determined by the power $a$ in eq.\ (\ref{spec}). In particular, it
is shown that for $a \geq 4$, the energy density itself remains finite
at $T_c$, only higher derivatives diverge there. The general structure
of the resulting thermodynamics near $T_c$ is thus that of critical 
behavior, associated with a phase transition leading to a new state of
matter, which we today take to be the quark-gluon plasma (QGP)
\cite{C-P}.

\medskip

The aim of this work is to determine in some detail the onset of
the critical behavior in strongly interacting matter, when coming from 
the confined hadronic side. The system here is an interacting hadron
system; but it is known that if the interaction is resonance
dominated, the medium of interacting elementary particles can be
replaced by an ideal gas of all possible resonances \cite{B-U,DMB}.
If we consider systems of vanishing overall baryon number density,
the assumption of resonance dominance appears to be well satisfied;
only for high baryon density do non-resonant repulsive forces come
into play \cite{CRS}. 

\medskip

We shall therefore consider here an ideal resonance gas of vanishing
overall baryon density, and study in particular the temperature 
dependence of the speed of sound, $c_s(T)$, in such a medium. By
definition,
\be
 c^2_s(T)~=~\bigg(\frac{{\partial}P}{{\partial}\varepsilon}\bigg)_V
~=~\frac{s(T)}{C_V(T)}, 
\label{soundspeed}
\end{equation}
where $s=(\partial P/\partial T)_V$ is the entropy density and 
\begin{equation}
 C_V(T)~=~\bigg(\frac{{\partial}\varepsilon}{{\partial}T}\bigg)_V.
\label{specheat}
\end{equation}
is the specific heat at constant volume. If the resonance gas has a 
suitable degeneracy structure of the Hagedorn/DRM type, the 
specific heat diverges at the limiting temperature $T_c$, while the 
entropy density remains finite there. As a result, the speed of sound
will vanish at $T_c$. It is thus expected that $c_s(T)$ is at low 
temperature $T$ that of an ideal gas of ground state hadrons (``pions''), 
until $T$ becomes high enough to bring resonance effects into play 
and eventually drive the system towards critical behavior. 

\medskip

The speed of sound has long been considered as a sensitive indicator
of critical behavior in strongly interacting matter. Early lattice
studies indeed found the expected sharp dip of $c_s^2$ in the
critical region \cite{GG,KR-HS}. Later, more precise studies seem to
indicate a weaker dip \cite{laterlattice}, in accord with the 
expectation that the deconfinement transition is a rapid cross-over, 
rather than a genuine phase transition with singular behavior of thermal
observables. The success of physical resonance gas models in the 
prediction of hadron abundances \cite{abundance} provided the stimulus
to also calculate the speed of sound in such schemes, using a 
grand canonical partition function of an ideal gas of all 
experimentally observed states up to a certain large mass.
Again one finds a decrease of $c_s^2(T)$ as the temperature is 
brought near the expected critical point \cite{laterlattice,physresgas}.
 
\medskip

In the next section we will evaluate the thermodynamical functions, 
first for an ideal gas of pions and then for an ideal resonance gas
with an exponentially growing mass spectrum of the form (\ref{spec}).
We shall refer to such a system as a Hagedorn resonance gas.
Following this, we calculate the resulting speed of sound and study
in detail how is approaches the expected critical behavior. 
In the final section, we compare our results to those obtained
for the physical resonance gas based on the actually observed resonances 
up to a mass limit of 2.5 GeV \cite{PDG}, as well as to the results from 
studies in finite temperature lattice QCD.

\section{Thermodynamics of a Hagedorn Gas}

For an ideal Boltzmann gas of identical scalar particles of mass 
$m_0$  and three charge states (``pions'') contained in a volume $V$, 
the grand partition function is defined as
\begin{equation}
Z(T,V)~=~{{\sum}_N}{\frac{1}{N!}}{\Bigg[\frac{3~V}{(2\pi)^3}{\int}d^3p
\exp\{-\sqrt{p^2 + {m_0}^2}/T\}\Bigg]}^{N}.
\label{piongas}
\end{equation}
This expression can be evaluated, giving
\begin{equation}
\ln Z(T,V)~=~3{\frac{VT{m_0}^2}{2{\pi}^2}}K_2(m_0/T),
\label{logpart}
\end{equation}
where $K_2(x)$ is the modified Bessel function of the second kind.
From this equation we obtain
\begin{equation}
P_0(T)~=~\left({\partial \ln Z \over \partial V}\right)_T~=~ 
{\frac{{3 m_0}^2{T^2}}{2{\pi}^2}}K_2(m_0/T),
\label{pressideal}
\end{equation}
for the pressure and
\begin{equation}
\varepsilon_0(T) ~= ~T^2 \left({\partial \ln Z \over \partial T}\right)_V~=
{3 m_0^2T^2 \over 2\pi^2}\left[3~ K_2(m_0/T)~+~
\left({m_0\over T}\right) K_1(m_0/T)\right]
\label{endenideal}
\end{equation}
for the energy density. Using these relations, 
\begin{equation}
s_0(T) ~=~{\varepsilon_0(T) + P_0(T)\over T}~=~
{3 m_0^2T \over 2\pi^2}\left[4~ K_2(m_0/T)~+~
\left({m_0\over T}\right) K_1(m_0/T)\right]
\label{entropdenideal}
\end{equation}
provides the entropy density.

\medskip

We now extend this to an ideal Boltzmann gas of resonances,
described by an exponentially increasing mass spectrum of the 
Hagedorn form (\ref{spec}). For the full spectrum of 
hadrons, including ground state pions as well the resonances,
we thus write
\begin{equation}
\rho(m)~=3 \delta(m-m_0)~ + ~Am^{-4}\exp\{m/T_c\}~\theta(m-2m_0). 
\label{massspectrum}
\end{equation}
Here the constant $A$ (of dimension $m^3$) provides the normalization of 
the resonance contributions relative to that of the pions, and
we have chosen the power term $a=4$ in order to obtain second
order singular behavior, as will be shown shortly. The $\theta$-function
assures that the resonance spectrum starts above the two-pion
threshold. With the spectrum (\ref{massspectrum}), the logarithm
of the grand canonical partition function is now given by
\begin{equation}
\ln Z(T,V)~=~{VT\over 2\pi^2}\left\{3 m_0^2 K_2(m_0/T)~+~
A{\int}^{\infty}_{\!2m_0}dm~m^{-2}~\exp\{m/T_c\}~K_2(m/T)\right\};
\label{logpartmass}
\end{equation}
the pressure thus becomes
\begin{equation}
P(T)~=~P_0(T)~+~
{\frac{A{T^2}}{2{\pi}^2}}{\int}^{\infty}_{2m_0}dm~{m}^{-2}~
\exp\{m/T_c\}~K_2(m/T),
\label{pressmass}
\end{equation}
with the pion gas pressure $P_0(T)$ given by eq.\ (\ref{pressideal}).
Similarly, we obtain  
\begin{equation}
\varepsilon(T) ~=~\varepsilon_0(T)~+~
{\frac{A{T^2}}{2{\pi}^2}}{\int}^{\infty}_{2m_0}dm~ {m}^{-2}~
\exp\{m/T_c\}~\left[3~K_2(m/T)~+ \left({m\over T}\right)K_1(m/T)\right],
\label{endenmass}
\end{equation}
for the energy density and
\begin{equation}
s(T)~=~s_0(T)~+~
{\frac{AT}{2{\pi}^2}}{\int}^{\infty}_{2m_0}dm~ {m}^{-2}~
\exp\{m/T_c\}\left[4~K_2(m/T)~+~\left({m \over T}\right)K_1(m/T)\right].
\label{entropmass}
\end{equation}
for the entropy density; here $\varepsilon_0(T)$ and $s_0(T)$ are
the corresponding pion gas quantities (\ref{endenideal}) and
(\ref{entropdenideal}).

\medskip

The Bessel functions appearing in these relations have the asymptotic 
form for large argument
\be
K_n(m/T) = \sqrt{\pi~T \over 2~m}~\exp\{-m/T\}~\left[1 + O(T/m)\right].
\label{Bessel}
\ee
This shows that the integrals determining pressure, energy and entropy
density all converge for for $T\leq T_c$, while diverging for $T>T_c$.
The convergence of the thermodynamics potentials at $T=T_c$ is
a direct consequence of our choice $a=4$ in the resonance spectrum;
for $a \leq 7/2$, both energy and entropy density would diverge there.
In our case, one further temperature derivative is needed to cause
a divergence, leading to a diverging specific heat at $T_c$. We shall 
return to the detailed form shortly.

\medskip

\begin{figure}[htb]
\centerline{\epsfig{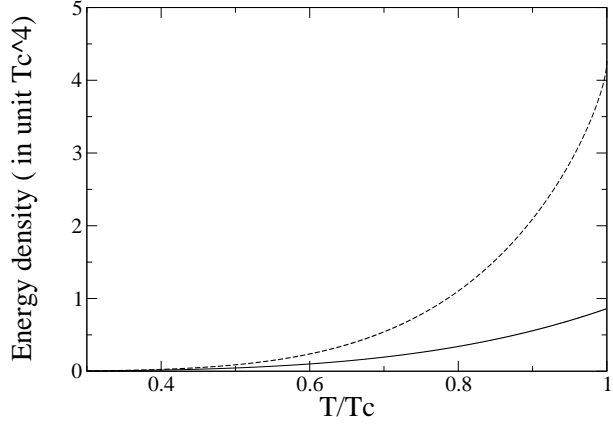}}
\caption{Energy density vs.\ temperature for an ideal pion gas
(solid lower curve) and an ideal resonance gas (dashed upper curve)}
\label{edens}
\end{figure}

Given the analytical form (\ref{massspectrum}) of the resonance spectrum,
both $\varepsilon_0(T)$ and $\varepsilon(T)$ can be evaluated numerically.
For this, the resonance weight $A$ has to be specified. With 
$\varepsilon(T_c)/T_c^4 \simeq 5$, as obtained in lattice studies for the 
case of two quark flavors \cite{KLP}, we obtain $A=15.35 ~T_c^3$. 
Using this value, we get the energy density behavior of the ideal 
resonance gas shown in Fig.\ \ref{edens}, where it is compared to that 
of the ideal pion gas. We see that around $T/T_c \simeq 0.6$, resonances 
come significantly into play, so that $\varepsilon(T)$ begins to increase 
above the pion gas value, until at $T_c$, resonances provide the
dominant part of the energy density.

\vskip0.5cm

\section{The Speed of Sound in a Hagedorn Gas}

\medskip

\begin{figure}[htb]
\vskip0.3cm
\centerline{\epsfig{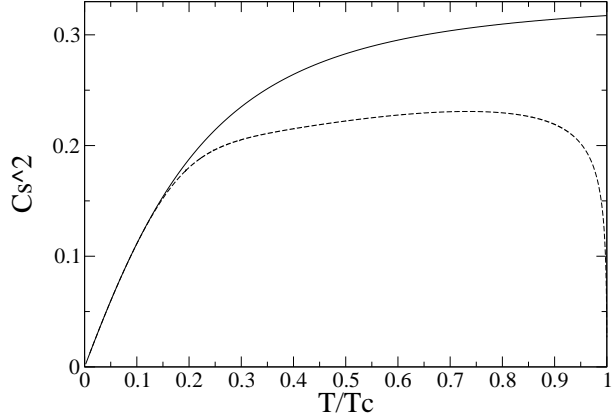}}
\caption{Speed of sound vs.\ temperature for an ideal pion gas 
(solid upper curve)
and a Hagedorn gas (dashed lower curve)}
\label{speed}
\end{figure}

The speed of sound $c_s$ at constant volume is defined by eq.\ 
(\ref{soundspeed}). For the case of the ideal pion gas we obtain
from the eqs.\ (\ref{specheat}) and (\ref{endenmass}) 
\begin{equation}
 C_V^0(T)~=~3s_0(T)~+~\frac{{3m_0}^4}{2{\pi}^2T }~K_2(m_0/T). 
\label{specheatpion}
\end{equation}
Together with the eqs.\ (\ref{soundspeed}) and (\ref{entropdenideal}) 
this leads to
\begin{equation}
\frac{1}{c^2_s}-3~={3 m_0^4 K_2(m_0/T) \over 2\pi^2 T s_0} ~=~
\frac{{m_0}^2K_2(m_0/T)}{4T^2K_2(m_0/T)+m_0TK_1(m_0/T)} 
\label{soundspeedideal}
\end{equation}
for the ideal pion gas. Making use of the large argument limit (\ref{Bessel})
of the Bessel functions, we find for $T \to 0$, the squared speed of sound 
$c_s^2$ vanishes linearly with $T$. The small argument limit
\be
K_n(x) ~=~ {2^{n-1}(n-1)! \over x^n} \left[1 -O(x^2)\right],
\label{Besselzero}
\ee
shows that for $T \to \infty$, we get $c_s^2 \to 1/3$. The overall
behavior below $T_c$ is shown in Fig.\ \ref{speed}.
 
\medskip

For the Hagedorn gas, the same line of argument gives
\begin{equation}
 C_V(T)~=~3s(T)~+~{1\over 2\pi^2 T}\left[3 m_0^4 K_2(m_0/T) ~+~
A {\int}^{\infty}_{2m_0}dm~\exp\{m/T_c\}~K_2(m/T)\right] 
\label{specheatres}
\end{equation}
for the specific heat. As $T\to T_c$, the entropy density remains
finite. However, the integral in eq.\ (\ref{specheatres}) diverges so that
at the Hagedorn temperature we have the critical behavior given as
\be
C_V(T) \sim (T_c - T)^{-\alpha},
\label{exp}
\ee
whereby the resulting critical exponent was found to be $\alpha=1/2$ \cite{S-ex}.  

\medskip

An immediate further consequence of the diverging specific heat is
that now the speed of sound vanishes at $T_c$. To obtain the functional
form of this behavior, we make use of eqs.\ (\ref{entropmass}), 
(\ref{soundspeed}) and (\ref{specheatres}) to get
\begin{equation} 
\frac{1}{c^2_s}~-3~={3 m_0^4K_2(m_0/T)~+~
A \int^{\infty}_{2m_0}dm~\exp\{m/T_c\}~K_2(m/T)
\over
2\pi^2 T s_0~+~
A~T{\int}^{\infty}_{2m_0}dm~ {m}^{-2}~\exp\{m/T_c\}~\left[4T K_2(m/T)~+~
m K_1(m/T)\right]} 
\label{soundspeedres}
\end{equation}
for the behavior ${c^2_s}$ in the Hagedorn gas. In this relation, the
second term in the numerator diverges as $T\to T_c$, which in turn
causes the speed of sound to vanish there.

\medskip

\begin{figure}[htb]
\centerline{\epsfig{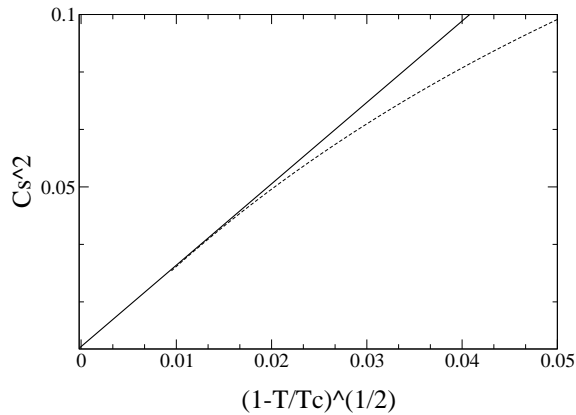}}
\caption{Speed of sound of a Hagedorn gas (dashed lower curve)
in the critical region, compared to the predicted critical behavior 
$(T_c-T)^{1/2}$ (solid upper curve)}
\label{crit}
\end{figure}

Making use of the value of $A$ obtained above, we show in Fig.\ \ref{speed}
the Hagedorn gas result compared to the ideal pion gas form.
It is seen that the Hagedorn gas follows the pion gas form closely 
until about $T/T_c \simeq 0.2$; from then on, any further increase
of the average energy density is mainly turned into resonance masses,
keeping the speed of sound roughly constant but non-zero. Only for 
temperatures very close to the critical point, for $T/T_c \geq 0.9$, 
do we reach a regime in which more and more of the energy goes into 
forming massive resonances. The truly critical region of the Hagedorn 
gas is thus restricted to a rather narrow band below the critical 
temperature. Magnifying this region, we show in  Fig.\ \ref{crit} 
the behavior of $c_s^2$ as function of the expected critical form 
$(T_c-T)^{1/2}$. It is seen that only in the last 1 - 2 \% below $T_c$
the predicted behavior actually occurs.

\vskip0.5cm

\medskip

\begin{figure}[htb]
\centerline{\epsfig{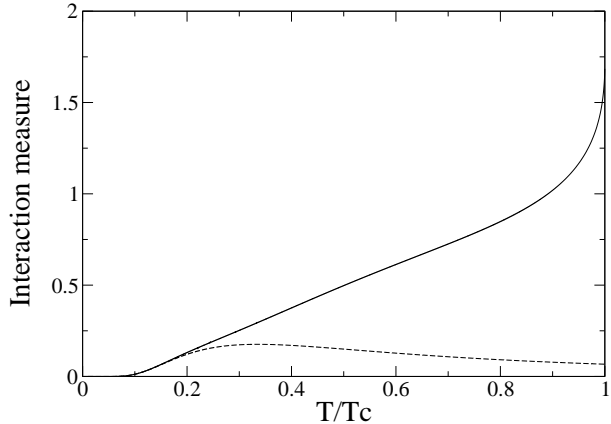}}
\caption{Interaction measure for an ideal pion gas (bottom curve) compared
to that of a Hagedorn gas (top curve)}
\label{inter} 
\end{figure}

We can obtain further information on the extent of the critical region
by making use of the interaction measure
\be
\Delta(T) \equiv {\e(T) - 3P(T) \over T^4},
\label{intermeas}
\ee
formed from the trace of the energy-momentum tensor; it vanishes for an
ideal gas of massless constituents. In Fig.\ \ref{inter}, we show the
modifications coming in for a resonance gas. Again we notice a rather
large region of ``normal'' resonance effects, until very close to $T_c$,
the critical rise sets in leading to a diverging slope of $\Delta(T)$
as $T \to T_c$. 

\vskip0.5cm

\section{The Low Mass Resonance Spectrum}

\begin{figure}[htb]
\centerline{\epsfig{file=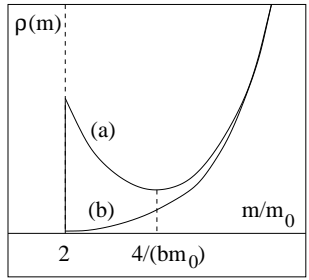,width=4cm}}
\caption{
Resonance spectrum for eq.\ (\ref{spec}) with $a=4$, compared to the
form given by eq.\ (\ref{thresh}).}
\label{dip}
\end{figure}

In this section we want to consider in some more detail the
effect of the resonance spectrum form in the low mass range on the
speed of sound $c_s(T)$. 
Our choice of the power factor of $\rho(m)$ in eq.\ (\ref{spec}) was dictated by
the desired behavior of the thermodynamic observables at the divergence
point $T_c$. When $a=4$, we had a finite energy density and a singular
specific heat. The crucial feature for this was, however, the behavior
at large $m$, for which a change in the threshold form at $m=2m_0$ would not
modify that property. The functional form for the spectrum (\ref{spec}) is shown
in Fig.\ \ref{dip}. It has a minimum at $m=4/b$, before it begins its
exponential rise. Thus it is evident that this minimum has no real physical
meaning and can be replaced by a more reasonable threshold behavior of the 
resonance contributions. Therefore, we can replace $\rho(m)$ in (\ref{spec}) by
\be
\rho_s(m) = \left[ 1 - \left({2m_0 \over m}\right)^s\right]m^{-4}
\exp\{bm\}
\label{thresh}
\ee
with some integer constant power $s>0$. Clearly $\rho_s(m)$ still retains 
the same critical behavior. The spectral form produced by (\ref{thresh})
for $s=2$ is included in Fig.\ \ref{dip}. Since the region of the low
mass resonances contributes very little to the functional form of
the speed of sound, the changes that such a threshold factor produces 
in $c_s^2(T)$ are effectively negligible.

\medskip

\medskip

\section{The Physical Resonance Gas}

We now want to compare our results to those obtained in studies of an 
ideal (Boltzmann) gas of actual physical resonances. The relevant partition 
function thus becomes the sum over all states listed by the 
Particle Data Compilation \cite{PDG} 
up to some mass value of $M$, above which the experimental information is
too incomplete to be of use. Typically one has $M = 2 - 3$ GeV. For each 
state of mass $m_i$, we include a degeneracy $g_i$ determined by the spin, 
isospin, baryon number and strangeness degrees of freedom.
The partition function is then given by
\be
\ln~Z_P(T,V) = \sum_{\rm i} \ln~Z^i_P(T,V),
\label{1}
\ee              
with
\be
\ln~Z_P^i(T,V) = {g_i \over (2\pi)^2} \int d^3 p \exp\left \{- 
{\sqrt{p^2 + m_i^2} \over T} \right\} = 
g_i~ {VT \over2 \pi^2}~ m_i^2 K_2(m_i/T),
\label{2}
\ee 
specifying the contribution of a resonance of $m_i$.

\medskip

\begin{figure}[htb]
\vspace*{0.5cm}
\centerline{\epsfig{file=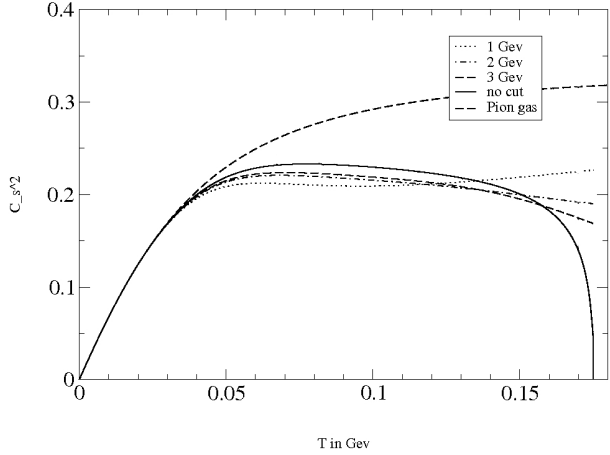,width=8cm}}
\caption{Speed of sound in a Hagedorn gas with resonance mass truncation,
$m \leq M$.}
\label{trunc}
\end{figure}

Before we calculate the speed of sound for such a ``physical'' resonance
gas, let us see what effect a mass truncation in the resonance integration
has for the Hagedorn gas. We thus consider the form obtained if we replace
in eq.\ (\ref{logpartmass}) the integral over $[2m_0,\infty]$ by one over
$[2m_0,M]$. The results are shown in Fig.\ \ref{trunc} for several values 
of $M$. Since the truncated system has no divergence at $T=T_c$, that is
since the specific heat now remains finite there, we can continue the 
$c_s^2$ curves also to values above $T_c$. Eventually, for any finite 
$M$, they will converge to $c_s^2=1/3$.

\medskip

\begin{figure}
\vspace*{0.5cm}
\centerline{\epsfig{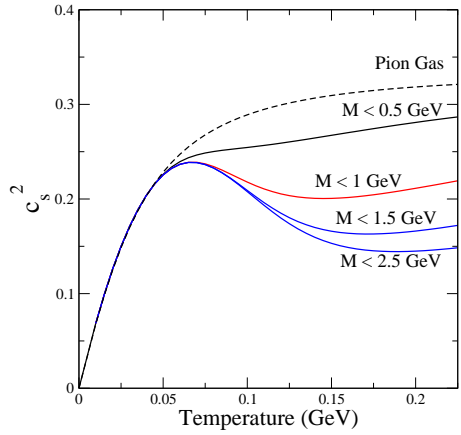}}
\caption{Speed of sound in a physical resonance gas with upper
limits $M$ in resonance mass.}
\label{cley-trunc}
\end{figure}

We now turn to the physical resonance gas and calculate the speed of 
sound for several values of the upper limit $M$ in resonance mass. The
results are shown in Fig.\ \ref{cley-trunc}. Qualitatively, they
agree quite well to what was found for the truncated Hagedorn gas.
It is seen that in both scenarios, the inclusion of further heavy 
resonances somewhat lowers the speed of sound in the temperature range 
$0 \leq T \leq 200$ MeV, but it never produces a sharp minimum. 

\medskip

\begin{figure}[htb]
\vspace*{0.5cm}
\centerline{\epsfig{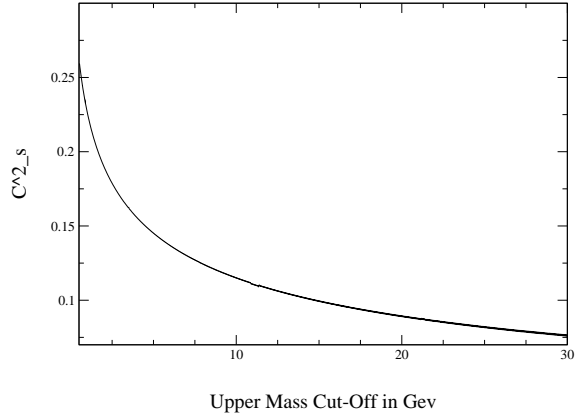}}
\caption{Speed of sound for a Hagedorn gas as function of the
upper limit $M$ in resonance mass.}
\label{masscut}
\end{figure}

Futher inspection does show, however, a difference in the behavior of
the two formulations. In Fig.\ \ref{masscut}, we show the speed of sound 
at $T=T_c$ in the Hagedorn gas for increasing resonance mass bounds. It
is seen to converge to zero for $M\to \infty$, which means that it is 
only the mass bound which prevents critical behavior. 

\medskip

\begin{figure}[htb]
\vspace*{0.5cm}
\centerline{\epsfig{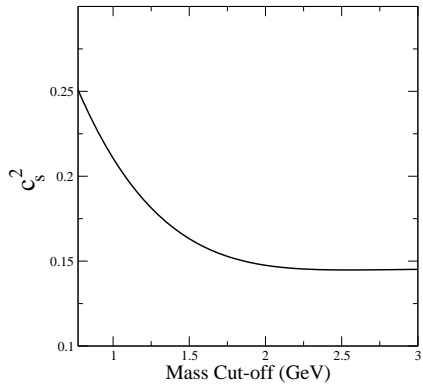}}
\caption{Speed of sound at $T=175$ MeV in a physical resonance gas, 
as function of upper resonance mass limit $M$.}
\label{masscut-cley}
\end{figure}

In the case of the physical resonance gas, we also have to consider 
the form of the degeneracy. In the degeneracy factor $g_i$, only the 
spin variable $J_i$ can result in an unbounded increase with $i$. 
For resonances on linear Regge trajectories, we have
\be
J_i \sim \alpha' m_i^2, 
\label{spin}
\ee
where $\alpha' \simeq 1$ GeV$^{-2}$ is the Regge resonance slope. Hence
the spin degeneracy $(2J_i+1)$ implies 
\be
g(m_i) \sim m_i^2.
\ee
It is clear that with such a degeneracy, the partition function will
remain analytic for all values of $T$, so that we will never encounter
critical behavior, no matter what the upper bound on the mass is. This
explains why, as seen in Fig\ \ref{masscut-cley}, an increase of
$M$ appears to produce a convergent finite value of $c_s^2(T_c)$. 

\medskip

Thus it is not only the upper limit in resonance mass which prevents the
physical resonance gas from ever showing critical behavior. The crucial 
factor is the use of a degeneracy determined by quantum number degrees of 
freedom only. Such a degeneracy grows, as we saw above, as $m^2$, and as a 
consequence, the partition function always remains analytic. The critical
behavior arising for a Hagedorn resonance gas is due to a degeneracy
growing exponentially in $m$. This, in turn, is a consequence of
counting as distinct degenerate states all partitions of a given
resonance into all other possible resonances. As we noted above, 
both the SBM and the DRM obtain the exponential mass degeneracy 
from the corresponding partition problem.

\medskip

Do these considerations have any physical implications? They clearly would,
if deconfinement were an actual phase transition with a diverging specific
heat. If it is only a rapid cross-over, or if the transition should lead to 
a finite specific heat at $T_c$ (such as produced in the $O(4)$ universality 
class), the behavior of the speed of sound obtained from the physical
resonance gas even at $T_c$ can remain a good approximation.   

\medskip

On the other hand, we have also seen that the actual critical region in
the Hagedorn gas is a very narrow band just below $T_c$. Up to a temperature
of $T\simeq 0.95~T_c$, the observed behavior is due to the normal
resonance pattern, and here physical resonance gas and Hagedorn gas
can certainly agree. This
leaves us with an amusing question: is the temperature obtained in
abundance analyses of hadron species indeed the temperature of the
confinement/deconfinement transition, or does the replacement of
an interacting hadron gas by a physical resonance gas break down very
close to a critical point?\\
~\\
{\bf{Acknowledgements}}
We would like to thank J\"urgen Engels and Olaf Kaczmarek for their help with
the evaluations.

\end{document}